\documentclass[12pt,longbibliography]{iopart}
\pdfoutput=1

\usepackage{graphicx}

\AtBeginDocument{%
    \newwrite\bibnotes
    \def\bibnotesext{Notes.bib}
    \immediate\openout\bibnotes=\jobname\bibnotesext
    \immediate\write\bibnotes{@CONTROL{REVTEX41Control}}
    \immediate\write\bibnotes{@CONTROL{%
    apsrev41Control,author="08",editor="1",pages="1",title="0",year="1"}}
     \if@filesw
     \immediate\write\@auxout{\string\citation{apsrev41Control}}%
    \fi
}%

\begin{document}

\title[Link translocation through a pore]{Translocation of links through a pore: effects of link complexity and size}

\author{M Caraglio$^1$, E Orlandini$^2$, S G Whittington$^3$}

\address{$^1$ Institut f\"ur Theoretische Physik, Universit\"at Innsbruck, Technikerstra{\ss}e  21A, A-6020 Innsbruck, Austria}
\address{$^2$ Dipartimento di Fisica e Astronomia ``Galileo Galilei", Sezione CNISM, Universit\`a degli Studi di Padova, via Marzolo 8, I-35131 Padova, Italy.}
\address{$^3$ Department of Chemistry, University of Toronto, Toronto M5S 3H6, Canada}
\ead{orlandini@pd.infn.it}
\vspace{10pt}

\begin{abstract}
We have used Langevin dynamics to simulate the forced translocation of linked polymer rings through a narrow pore.
For fixed size (\emph{i.e.} fixed number of monomers) the translocation time depends on the link type and on whether the rings are knotted or unknotted.
For links with two unknotted rings the crossings between the rings can slow down the translocation and are responsible for a delay as the crossings pass through the pore.
The results fall on a set of relatively smooth curves for different link families with the translocation time not always increasing with crossings number within the same family.
When one ring is knotted the results depend on whether the link is prime or composite and, for the composite case, they depend on whether the knotted or unknotted ring enters the pore first. 
We find a similar situation for 3-component links where the results depend on whether the link is prime or composite.
These results contribute to our understanding of how the entanglement complexity  between filaments impacts their translocation dynamics and should be useful for extending nanopore-sensing techniques to probe the topological properties of these systems.
\end{abstract}

\vspace{2pc}
\maketitle

\section{Introduction}
There are interesting questions about the behaviour of polymers when they translocate
through a narrow pore or orifice~\cite{Palyulin_et_al_SoftMatter_2014}.
In particular, there are several important scenarios in biological situations.
These include the ejection of DNA molecules from viral capsids~\cite{Kindt_et_al_PNAS_2001,Marenduzzo_et_al_PNAS_2009,Matthews:2009:Phys-Rev-Lett:19257792,Marenduzzo_et_al_PNAS_2013} and the transport of biopolymers such as proteins across cell membranes~\cite{Schatz_Dobberstein_1996,Nakielny_Dreyfuss_Cell_1999,Wickner_Schekman_Science_2005}.  
Driven translocation through solid state nanopores is a useful technique for probing various properties of polymers, including DNA sequencing~\cite{Zwolak_DiVentra:2008:RevModPhys,Kasianoviz:1996:PNAS}, detection of folded configurations~\cite{Huang_Makarov:2008:JCP,Haque_et_al_Biomaterials_2015} and mechanical unzipping of double-stranded DNA~\cite{Sauer-Budge_et_al_PRL_2003}.

It is well known that very long polymers are inevitably entangled~\cite{Orlandini_Whittington_RMP_2007}
This includes topological self-entanglement such as knotting and geometric self-entanglement such as writhe (or supercoiling).
Polymers can also be mutually entangled~\cite{Adams:1992:Cell,Orlandini:1994:J-Phys-A,Arsuaga:2007:J-Phys-A,Hirayama:2009:J-Phys-a,Soteros:2009:JKTR}.
For instance, ring polymers can be linked in various ways.
Since these entanglements are essentially always present for sufficiently long polymers the question arises as to how their presence affects the process of translocation through a pore.
This question has been studied for knotted rings and the knotted part of the ring translocates last, leading to an additional delay towards the end of the translocation process~\cite{Rosa:2012:PRL,Suma_et_al_MacroLett_2016}.  
This means that a translocation experiment can be used to detect knots in a ring polymer and might give information about the complexity of the knot~\cite{Plesa_et_al_NatNato_2016,Sharma2019,Suma_Micheletti_PNAS_2017}.
If instead we have a link with two components, both components being the same size and unknotted, the linked portion~\cite{Caraglio_et_al_ScieRep_2017} gives rise to an additional delay around the middle of the translocation process~\cite{Caraglio_et_al_macromol_2017}.  
For a link with three unknotted components (such as the connect sum of two Hopf links) there are two separate additional delays, around one third and two thirds of the way through the process~\cite{Caraglio_et_al_macromol_2017}.
This suggests that a translocation experiment could detect linking and might give information about the link type.
In this paper we investigate the translocation, through a pore, of various link types~\cite{Rolfsen:1976}.
Specifically, we look at cases where we have a link of two components, neither of which is knotted, and examine how the translocation depends on the link type and, specifically, on the complexity of the link. 
We also look at cases where one or both rings are knotted and where the link is prime (such as $7_{4}^2$) or composite (such as the connect sum of a Hopf link and a trefoil). 
We see quite different behaviour in the two cases.
Similarly, we look at 3-component links and contrast the behaviour of prime (such as $6_1^3$) and composite (such as $2_1^2\# 2_1^2$) links.
In addition, we investigate how the contour length  of the link affects the translocation time and how this depends on the typical size and on the definition of the linked portion.

\section{Model and simulation details}
\begin{figure*}
 \includegraphics[width=\textwidth]{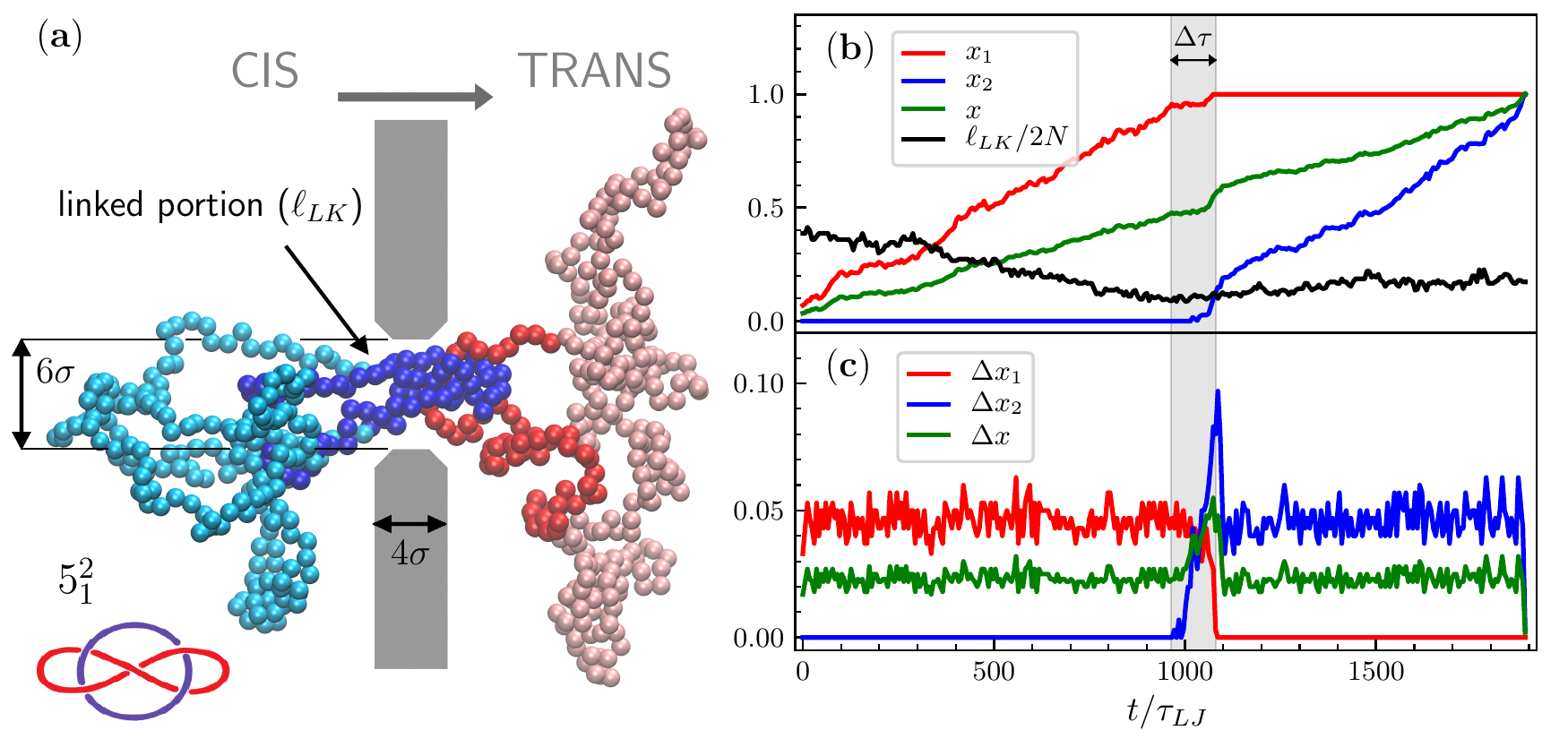}
 \caption{(a) A Whitehead link ($5^2_1$), between two rings of $N = 300$ beads 
 translocating through the pore. Highlighted parts of the rings represent the
 linked portion. (b) Time evolution of the translocated fraction $x$ (green curve) 
 and of the linked portion $\ell_{LK}$ (black curve). Red and blue curves refer to 
 the translocated fraction of ring 1 ($x_1$) and ring 2 ($x_2$), respectively. 
 (c) Time dependence of the fraction of beads inside the pore (colours as in panel b).
 The grey region highlights the time $\Delta \tau$ during which the translocation stalls.}
 \label{fig:model}
\end{figure*}

Linked loops are modelled as circular flexible chains of $N$ beads of diameter $\sigma$, connected by a {\small FENE} potential while a Weeks-Chandler-Andersen potential of energy $\epsilon = k_B T$ provides excluded volume interactions which maintain the topology~\cite{FENE}.
The pore is modelled as an hourglass-shaped channel composed of an inner cylinder of length $2\sigma$ and radius $3\sigma$, and two truncated cones of length $\sigma$ and outer radius $4\sigma$ drilled through an impenetrable wall of thickness $4\sigma$ that divides the space into the CIS (i.e., where the links start) and the TRANS region (see Figure~\ref{fig:model}a).
Since we are focusing only on the translocation process, we consider starting configurations obtained by threading a portion of one ring through the pore and let the whole system equilibrate keeping the beads inside the pore fixed.
Translocation starts by switching on a longitudinal electric field that exerts a local force $f$, directed towards the TRANS region, only on beads within the inner part of the pore.
In this study we keep fixed both the pore width, to $6\sigma$, and the driving force, to $f=0.5 k_BT/\sigma$. 
Constant-temperature Langevin dynamics is integrated with the {\small{LAMMPS}} package~\cite{LAMMPS} with standard values for the friction coefficient and bead mass resulting in a characteristic simulation time $\tau_{LJ} = \sigma \sqrt{m/k_B T}$.
For different topologies and various $N$ considered we performed $100$ independent trajectories that last until the translocation process is concluded, i.e. when all the beads are in the TRANS regions.

By following the position of the link with respect to the pore we have calculated several properties of the translocation process, including the translocated fraction $x$ of the link, the total translocation time $\tau$, the translocation time for the first ring $\tau_1$ and the stalling time $\Delta \tau$ defined as the time at which the first component has completely translocated minus the time at which the second component first enters the pore (see Figure~\ref{fig:model}b).
Another quantity that is often measured in translocation experiments is the so-called waiting time $w = dt/dx$, i.e., the inverse of the translocation velocity.

Finally, the translocation process should depend strongly on how the entangled portion of the link passes through the pore and, to monitor quantitatively this property, we have estimated the contour length $\ell_{LK}$ of the portion of the link involved in the physical link (linked portion) (see~\cite{Caraglio_et_al_ScieRep_2017,Caraglio_et_al_Polymers_2017,Amici_et_al_macrolett_2019}).

\section{\label{sec:level3} Dependence on link complexity }
In this section we describe our results on how the link translocation process depends on the topological complexity of the link (measured for instance by crossing number). 
We first consider two-components links that are either both unknotted or in which at least one is knotted. 
We next consider a few examples for the more complex case of three component links.
It will transpire that it is useful to separate the crossing number into two parts: the number of crossing between two rings and the number of self-crossings in a ring.

\subsection{Dependence on link complexity for 2-component links}

\begin{figure*}
 \centering
 \includegraphics[width=\textwidth]{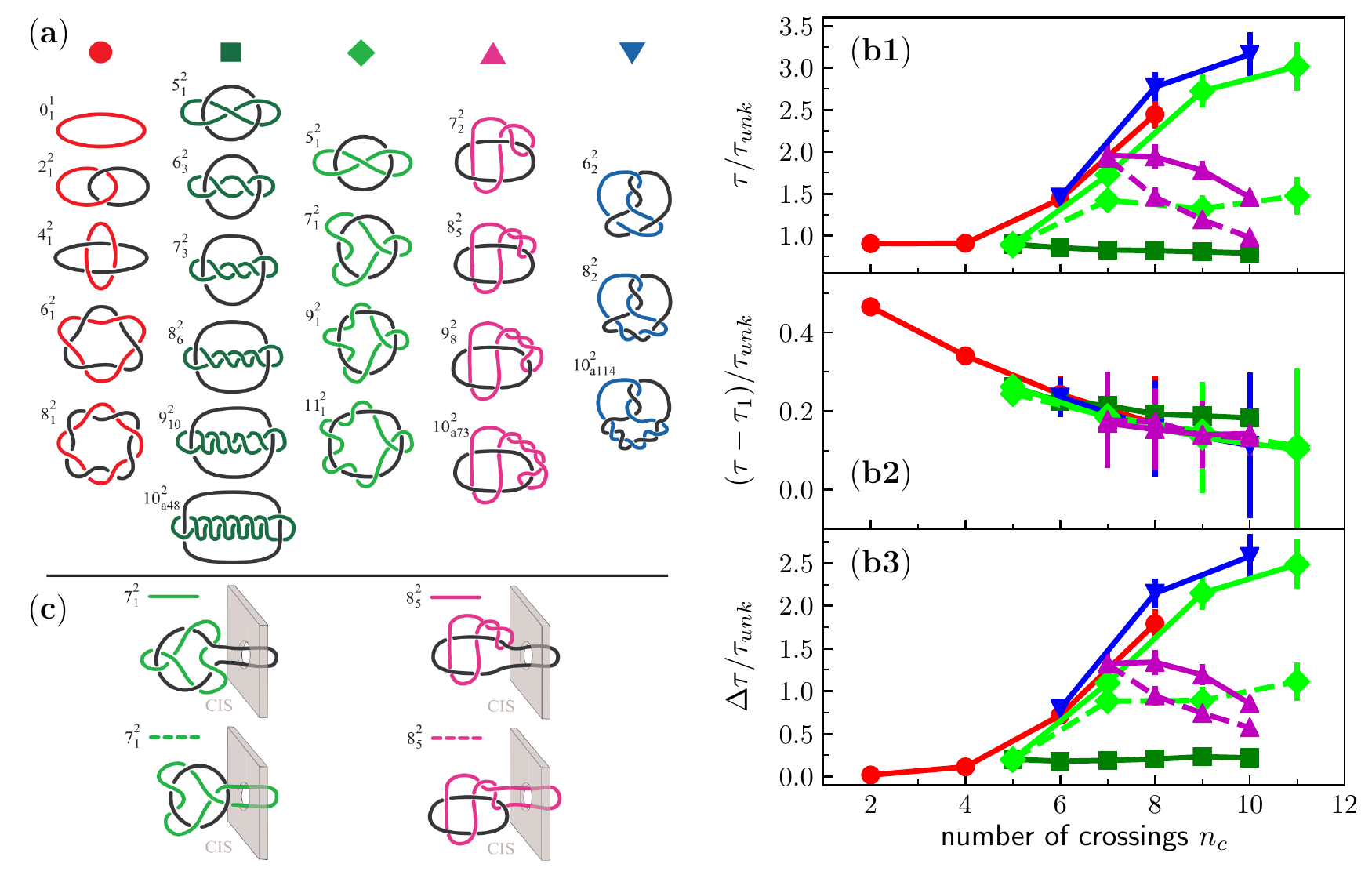}
 \caption{(a) Minimal crossing representation of the link types considered in this study.
 Each column (and colour) refers to links belonging to the same family.
 (b1) Total translocation time $\tau$, (b2) Time required for the second component to
 translocate  and (b3) The stalling time $\Delta \tau$. For comparison all these times 
 are divided by the average translocation time of a single unknotted  ring with the same
 length $2N$.  (c) Two examples of links whose translocation dynamics depends on which
 component enters the pore  first.}
 \label{fig:fig2}
\end{figure*}

By fixing the link size at $2N=600$ we study the driven translocation dynamics for the following families of links\footnote{We use the Rolfsen notation~\cite{Adams:1994}.
However, this notation does not deal with links having a minimal crossing number greater than 9 and in these cases we adopt the notation used in the Knot Atlas~\cite{KnotAtlas}}:
\begin{enumerate}
\item
the $(2,2k)$-torus links $2_1^2$, $4_1^2$, $6_1^2$, and $8_1^2$,
\item
the family $5_1^2$, $6_3^2$, $7_3^2$, $8_6^2$, $9_{10}^2$,
and $10_{a48}^2$;
\item
the family $5_1^2$, $7_1^2$, $9_1^2$, and $11_1^2$;
\item
the family $7_2^2$, $8_5^2$, $9_8^2$ and $10_{a73}^2$;
\item
the family $6^2_2$, $8^2_2$, $10^2_{a114}$;
\item
2-component prime links with one component knotted: $7_4^2$, $7_7^2$, $8_{10}^2$, and
$8_{12}^2$;
\item composite links $2_1^2 \# 3_1$, $2_1^2  \# 4_1$,  $4_1^2 \# 3_1$,
$4_1^2 \# 4_1$, etc. in which one of the components is knotted as either a trefoil ($3_1$) or a figure-eight ($4_1$) knot, etc.
\end{enumerate}
These links are shown in Figure~\ref{fig:fig2}a using their $2D$ minimal crossing representation.

\subsection{2-component links with unknotted components}

If we plot the translocation time against the crossing number of the link, for fixed link size, we see a scatter of points.
However, the various links fall into link families and then the translocation times fall on a family of relatively smooth curves.
In Figure~\ref{fig:fig2}b we show the dependence of $\tau$,  $\tau-\tau_1$ and $\Delta \tau$ on crossing number for the various link families.
Note that for comparison these times are divided by the average time $\tau_{un}$ needed for  a single (unknotted) ring of the same length  ($2N=600$) to translocate. 

The behaviour of $\Delta \tau/\tau_{un}$ largely follows the behaviour of the total translocation time, $\tau/\tau_{un}$, while $(\tau - \tau_1)/\tau_{un}$, the translocation time of the second link decreases with increasing crossing number and is largely independent of link type at fixed crossing number.
The decrease as crossing number increases is because there is less of the second component to translocate when the relevant crossings have passed through the pore.

For the $(2,2k)$-torus links we note that the Hopf link $2_1^2$ and the Solomon link $4_1^2$ both translocate faster than a single unknotted ring of the same size.
The two links are less maximally extended in the direction of the force than a single unknotted ring and the linked portion is not large enough to obstruct the pore sufficiently to offset this effect.
For larger numbers of crossings the translocation time increases steadily with increasing complexity.
$\tau$ is monotonically increasing with crossing number, as is $\Delta \tau$.
These quantities reflect the increasing effect of pore blocking as the complexity increases.

For the family $6^2_2$, $8^2_2$, $10^2_{a114}$ we have similar behaviour in that the translocation time $\tau$ increases steadily with increasing crossing number.
$\Delta \tau$ also increases monotonically.

At the other extreme we have the family $5_1^2$, $6_3^2$, $7_3^2$, etc. where the
translocation time depends very weakly on the crossing number and is a sightly decreasing
function of crossing number.
Here it is useful to distinguish between the inter-ring crossings (remaining constant at 4) and the intra-ring crossings (increasing steadily with complexity).
The intra-ring crossings seem to have little effect on the translocation process, except to decrease slightly the maximal extension.
They slightly decrease $\tau$, probably because the link becomes less extended in the translocation direction, but have essentially no effect on $\Delta \tau$.

For the $7_2^2$, $8_5^2$, $9_8^2$ family the translocation time is a \emph{decreasing} function of the crossing number.
The number of inter-ring crossings remains the same but the self-entanglement of
one ring (measured by the intra-ring crossings) increases.
This will decrease the extension in the translocation direction and this seems to outweigh the blocking effect of the extra crossings.

If we examine Figure~\ref{fig:fig2}(b1) we see a general trend when we fix the total crossing number. 
For instance, when the crossing number is 6, $\tau$ is largest for $6_1^2$ and for $6_2^2$ and decreases for $6_3^2$, reflecting the fact that the number of inter-circle crossings decreases from 6 ($6_1^2$ and $6_2^2$) to 4 ($6_3^2$).  
We see the same behaviour for $8_1^2$ and $8_2^2$ (8 inter-circle crossings), $8_5^2$ (6 inter-circle crossings) and $8_6^2$ (4 inter-circle crossings).
For larger crossing numbers the situation is more complicated since $\tau$ can depend on which component enters the pore first.
This is not surprising when one component is knotted and the other is unknotted (see later) but it is perhaps more surprising when both rings are unknotted.
For small crossing number the effect, if it exists, is very small, but it is more important for links with larger crossing number.
For instance, for $7_1^2$ and other members of the same family the translocation is slower when the ring with no intra-ring crossings enters first, compared to when the ring with intra-ring crossings enters first.  
We see the same effect for $8_5^2$ (see Fig.~\ref{fig:fig2}c). 
For the torus links this effect is absent since there is symmetry between the two components.
For the other links, although the two components can be interchanged by smooth deformations if neither has vertices fixed in the pore, this is not the case once a ring has started to enter the pore.
In an experimental situation, at least at low force, a ring could exit the pore on the CIS side and the other ring might enter first, but this is not allowed in our simulation.
It may be that the intra-ring crossings pass through more easily at the beginning of the translocation process than when they are involved with the inter-ring crossings later in the process.

\subsection{2-bridge links and 4-plats}
\begin{figure*}
 \centering
 \includegraphics[width=\textwidth]{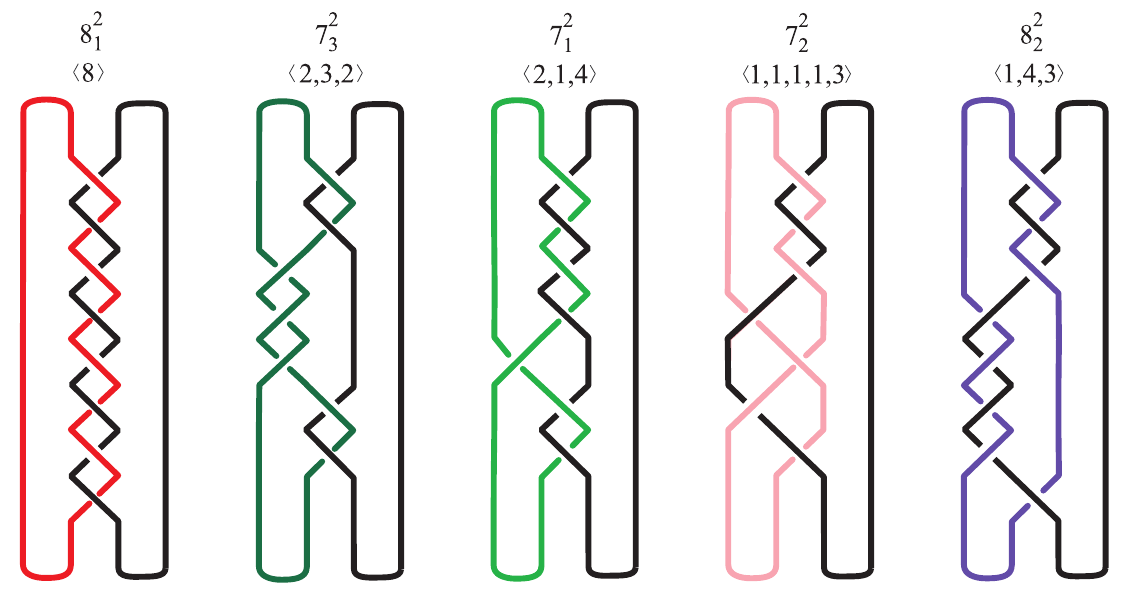}
 \caption{4-plat representations of some links, one for each family, considered in this
 study. 
}
 \label{fig:4plats}
\end{figure*}

All of the links discussed above are 2-bridge links and so they can be represented as 4-plats (see Figure~\ref{fig:4plats})~\cite{Burde_Zieschang}.   
This turns out to be a convenient way of representing each of the families that we have considered.
A 4-plat can be thought of as being derived from a braid on 4 strings where one of the strings has no crossings.
It can be characterized by a word in the braid group $B_3$, \emph{i.e.} as an alternating sequence of crossings of string 2 with string 3 and string 1 with string 2, so it is a sequence $ \sigma_2^{a_1} \sigma_1^{a_2} \sigma_2^{a_3} \ldots \sigma_2^{a_r}$ and the sequence $a_1,a_2, \ldots a_r$ forms the Conway symbol of the 4-plat. 
$\sum_k a_k$ is the crossing number of the link\cite{Burde_Zieschang}.  
We are not concerned with chirality so, for a chiral link, we can choose either of the two possible 4-plat representations.
The torus links $2_1^2$, $4_1^2$, $6_1^2$  etc. can be represented as $\langle 2p \rangle$ for $p=1,2,3, \ldots$.
Similarly, the family $5_1^2$, $6_3^2$, $7_3^2$ etc. can be represented as $\langle 2,p,2 \rangle$, for $p=1,2,3, \ldots$, the family $5_1^2$, $7_1^2$, $9_1^2$ etc. as $\langle 2,1,2p \rangle$, for $p=1,2,3, \ldots$, and the family $7_2^2$, $8_5^2$, etc. as $\langle 1,1,p,1,3 \rangle$ for $p=1,2,\ldots $.
The family $6_2^2$, $8_2^2$, $10_{a114}^2$ can be represented as $\langle 1,2p,3  \rangle$, $p=1,2,3, \dots$.
For the torus links $\langle 2p \rangle$, for the family $\langle 2,1,2p \rangle$ and for the family $\langle 1,2p,3 \rangle$, the number of inter-circle crossings is $2p$.
On the contrary, for the family $\langle 2,p,2 \rangle$ and for the family $\langle 1,1,p,1,3 \rangle$, $p$ counts the number of intra-circle crossings.

As we increase the value of $p$ each of these 4-plat families lies on a relatively smooth curve for each of the translocation properties that we have calculated. 
In we consider only the cases in which the component without intra-circle crossings enter first, for $\langle 2p \rangle$, $\langle 1,2p,3 \rangle$ and for $\langle 2,1,2p \rangle$ the translocation time increases smoothly with increasing $p$ while for $\langle 2,p,2 \rangle$ and for $\langle 1,1,p,1,3 \rangle$ there is less dependence on $p$ indicating that intra-circle crossings have a much smaller effect on the translocation time.
Their primary effect is to decrease the length in the translocation direction and this sometimes leads to a decrease in the translocation time.

\subsection{2-component links with one knotted component}
\begin{figure*}
 \centering
 \includegraphics[width=\textwidth]{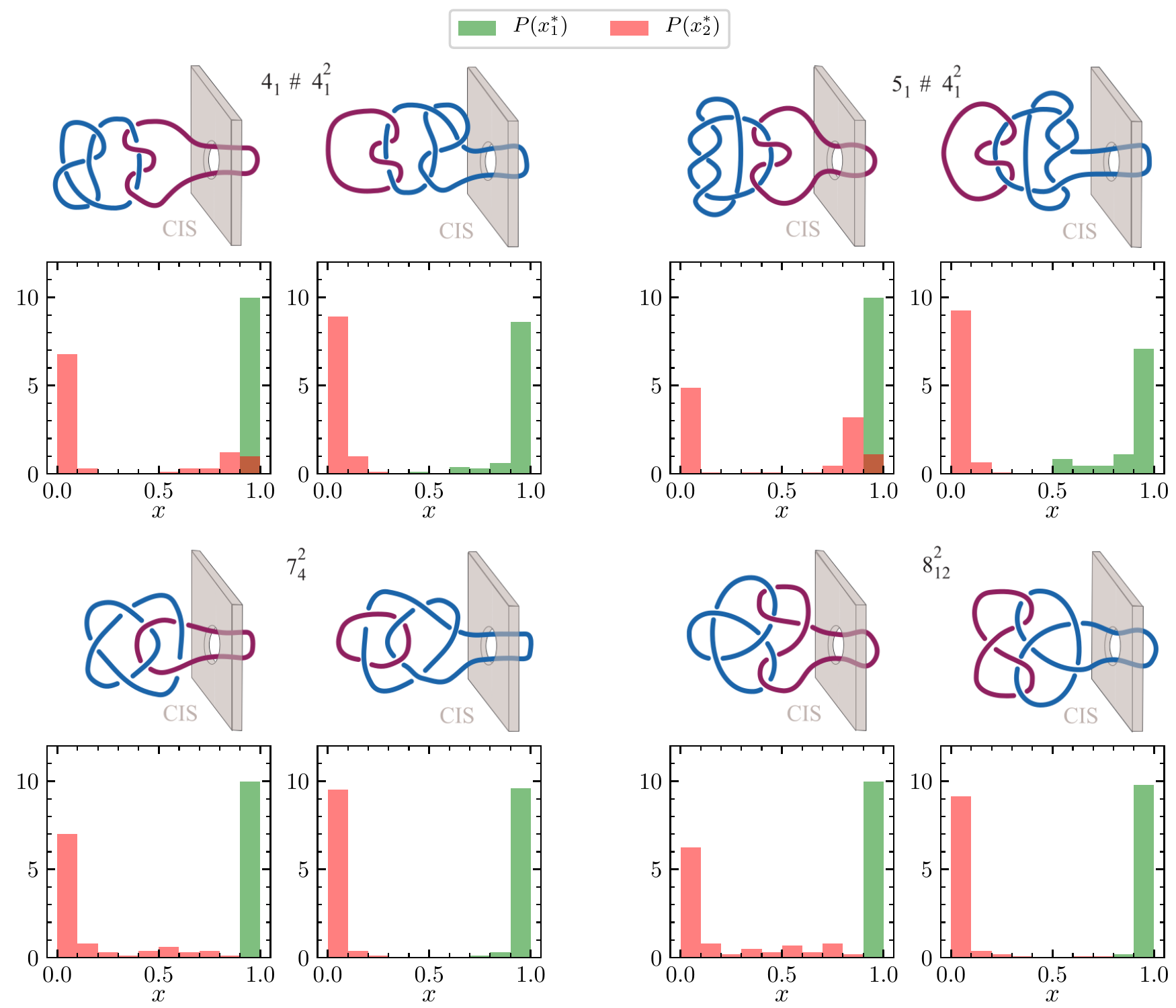}
 \caption{Distributions $P(x_1^*)$ and $P(x_2^*)$ (defined in the text) for links in
 which one component is knotted.
}
 \label{fig:fig_one_knotted}
\end{figure*}

When we consider cases where one component is knotted we have two different scenarios.
The link can be composite, \emph{e.g.} the connect sum of a prime link with both components unknotted, and a prime knot such as $3_1$ or $4_1$.
Alternatively the link can be prime with one knotted component, such as $7_4^2$ or $7_7^2$, where one component is knotted (a trefoil in these cases) but the link cannot be decomposed into a prime link with unknotted components, and a knot (see cartoons in Fig.~\ref{fig:fig_one_knotted}).  
We first consider the case of composite links with one component knotted.  
In each case that we have considered the general form is $L \# K$ where $L$ is a 2-component prime link with unknotted components (such as $2_1^2$ or $4_1^2$) and $K$ is a prime knot (such as $3_1$ or $4_1$).
We examine $P(x^*)$, the distribution of the fraction $x$ at which peaks in the signal of $\Delta x$ are observed (see Fig.~\ref{fig:model}c for an example), and the corresponding distributions $P(x_1^*)$ and  $P(x_2^*)$ for the two individual rings, where ring 1 enters the pore first (see Figure~\ref{fig:fig_one_knotted}). 
In general, the distributions depend on whether the knotted or unknotted ring enters the pore first.

Suppose that the unknotted component enters the pore first and suppose that $L$ is $4_1^2$.  
We have considered the cases where $K$ is $3_1$, $4_1$, $5_1$ and $7_1$. 
For each of these cases $P(x_1^*)$ has one major peak close to $x_1^*=1$ so that the primary obstruction occurs when most of the first component has passed through the pore. 
In this case, the distribution $P(x_1^*)$ does not help to understand if the main obstruction is due to the crossings between the two components or to the inter-circle crossings.  
On the other hand, the distribution $P(x_2^*)$ shows two peaks, one around $x_2^*=0$ and the second close to $x_2^*=1$.
The importance of the second peak increases as the complexity (\emph{i.e.} the crossing number) of $K$ increases.
The first peak comes from the passage of the crossings between the two rings and the second from the passage of the knotted part of the second ring.
If the knotted component enters the pore first the main peak in $P(x_1^*)$ is around $x_1^*=1$ and the main peak in $P(x_2^*)$ is around $x_2^*=0$, suggesting that all the crossings (both inter- and intra-crossings) pass through the pore more or less together.
For $3_1 \# 4_1^2 \# 3_1$ where both components are knotted (as trefoils) we have similar behaviour with a peak in $P(x_1^*)$ close to $x_1=1$ and peaks in $P(x_2^*)$ around $x_2^*=0$ and $x_2^*=1$ (not shown).
The intra-ring crossings in the second component enter the pore last.
By symmetry, this is independent of which component enters first.

As examples of prime links where one component is knotted we have investigated the behaviour of $7_4^2$, $7_7^2$, $8_{10}^2$ and $8_{12}^2$ (see Fig.~\ref{fig:fig_one_knotted} for the cases $7_4^2$ and $8_{12}^2$).  
In each case one component is a trefoil and the other component is unknotted.
In these cases the main peak in $P(x_1^*)$ is close to $x_1^*=1$ and the main peak in $P(x_2^*)$ is around $x_2^*=0$.
There is no peak in $P(x_2^*)$ around $x_2^*=1$, whichever component first enters the pore, indicating that all the crossings pass through close together and form a single obstruction.
This reflects the fact that the intra-ring crossings in the knotted component are intermingled with the inter-ring crossings, unlike the composite case where they are quite separate.
This is a clear distinction between the prime and composite cases.
For the prime cases there is also some additional structure around $x=1/2$.
This might be associated with the intra-ring crossings in the knotted ring.
These presumably pass through the pore after the inter-ring crossings.

\subsection{3-components links}
\begin{figure*}
 \centering
\includegraphics[width=\textwidth]{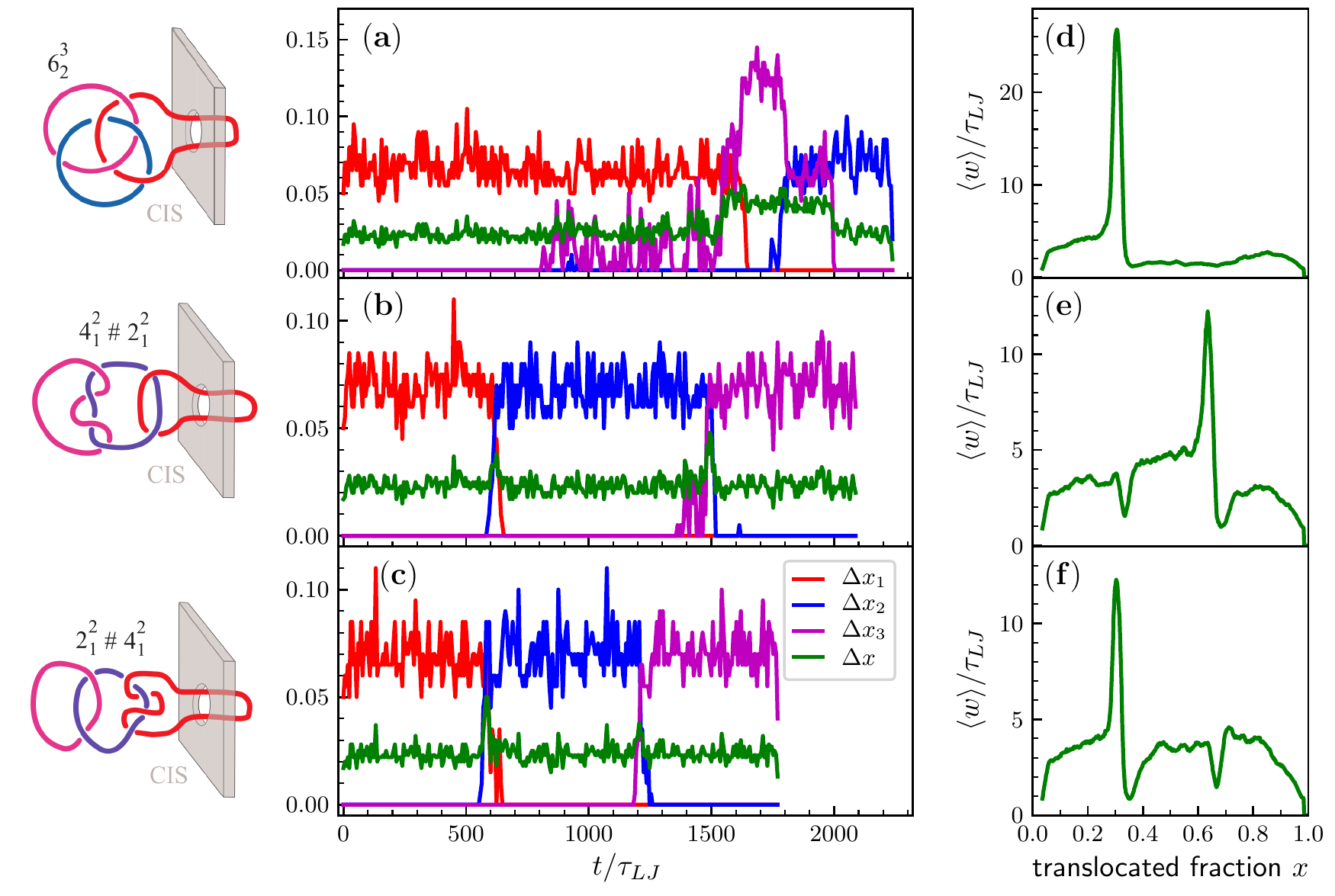}
 \caption{(a-c) Time evolution of the fraction of beads inside the pore for 1$^{st}$
 (red), 2$^{nd}$ (blue) and 3$^{rd}$ (magenta) components and of the total fraction of
 beads inside the pore (green) for respectively the Borromean link (a), 
 $4^2_1 \, \# \,2^2_1$ (b) and $2^2_1 \, \# \, 4^2_1$ (c). (e-f) waiting time as a
 function of the translocated fraction $x$.}
 \label{fig:trajectory3complinks}
\end{figure*}

\begin{figure*}
 \centering
 \includegraphics[width=\textwidth]{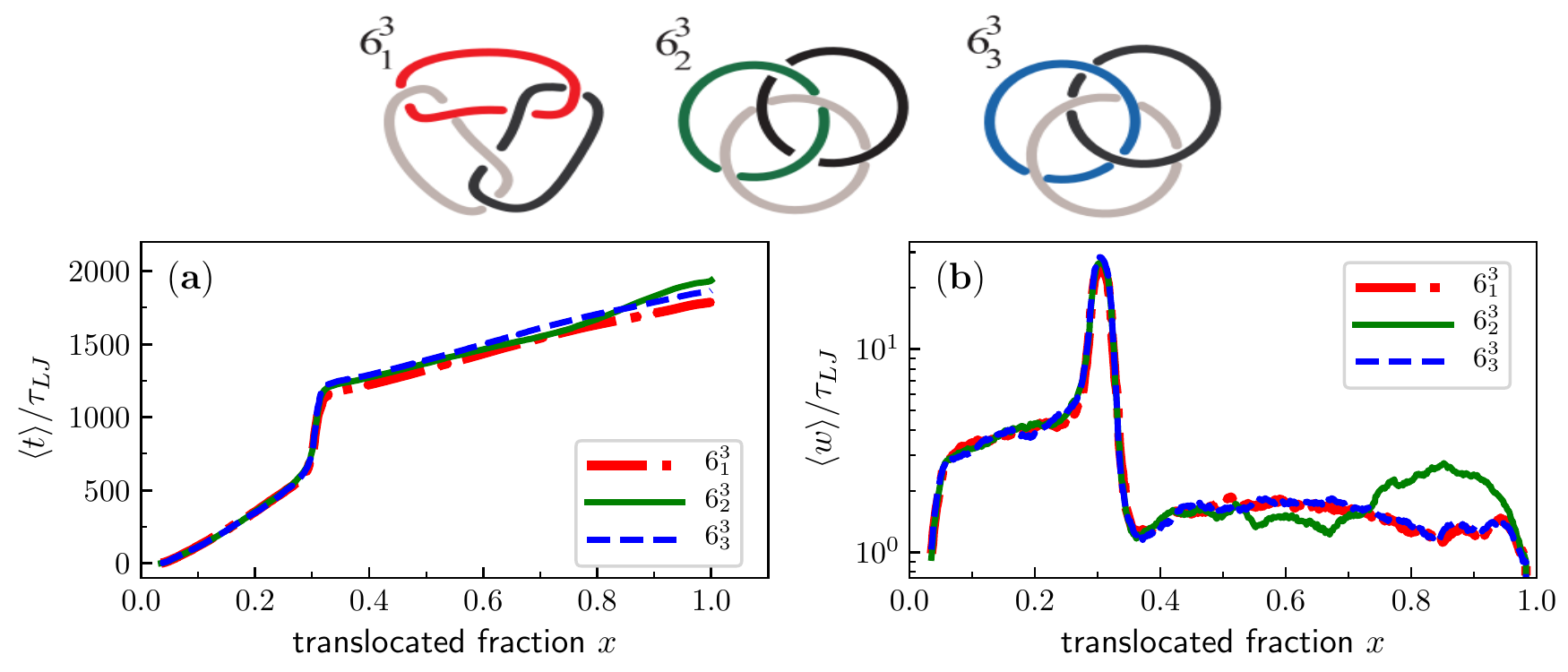}
 \caption{The three prime 3 components link show a very similar behaviour both in (a)
 the average time required to translocate a fraction $x$ of the total contour length and
 in (b) the waiting time as a function of the translocated fraction.}
 \label{fig:fig6}
\end{figure*}

We have also investigated the behaviour of 3-components links. 
In particular we focus on links with minimal crossing number equal to 6:
\begin{enumerate}
\item prime 3-component links $6^3_1$, $6^3_2$ (the Borromean link) and $6^3_3$.
\item a composite 3-component  link $2^2_1 \, \# \, 4^2_1$ in which the $4^2_1$ is the first to translocate and its counterpart $4^2_1 \, \# \, 2^2_1$ for which the $2^2_1$ translocates first.
\end{enumerate}
Figure~\ref{fig:trajectory3complinks}(a-c) shows the time dependence of the fraction $\Delta x$, $\Delta x_1$, $\Delta x_2$ and $\Delta x_3$ of beads at the pore respectively of the full link, the first, the second and the third component. 
One can see that the Borromean link ($6_2^3$) displays a single peak in the delay time indicating that all its crossings are passing through the pore within a short time interval.
A similar behaviour is observed for the other two prime 3-component links (see Figure~\ref{fig:fig6}b). 
For all three links we have a peak in $P(x_1^*)$ close to $x_1^*=1$ and peaks in $P(x_2^*)$ and $P(x_3^*)$ close to $x_2^*=0$ and $x_3^*=0$ (not shown): all the crossings pass through the pore at about the same time.

On the contrary, for the composite links we see two delays indicating that the crossings between the two pairs of circles pass through with a time delay in between (see Figure~\ref{fig:trajectory3complinks}(e,f)).
If the pair with more crossings passes first, the first waiting time is longer, and \emph{vice versa}.
$P(x_1^*)$ has a peak around $x_1^*=1$, $P(x_2^*)$ has two peaks, around $x_2^*=0$ and 1, and $P(x_3^*)$ has a single peak around $x_3^*=0$ (not shown).

The above results suggest that by measuring the waiting time, a quantity accessible in translocation experiments, one should be able to distinguish prime and composite 3-component links.

\section{\label{sec:level4} Effect of link size on translocation}

\begin{figure*}
 \centering
 \includegraphics[width=\textwidth]{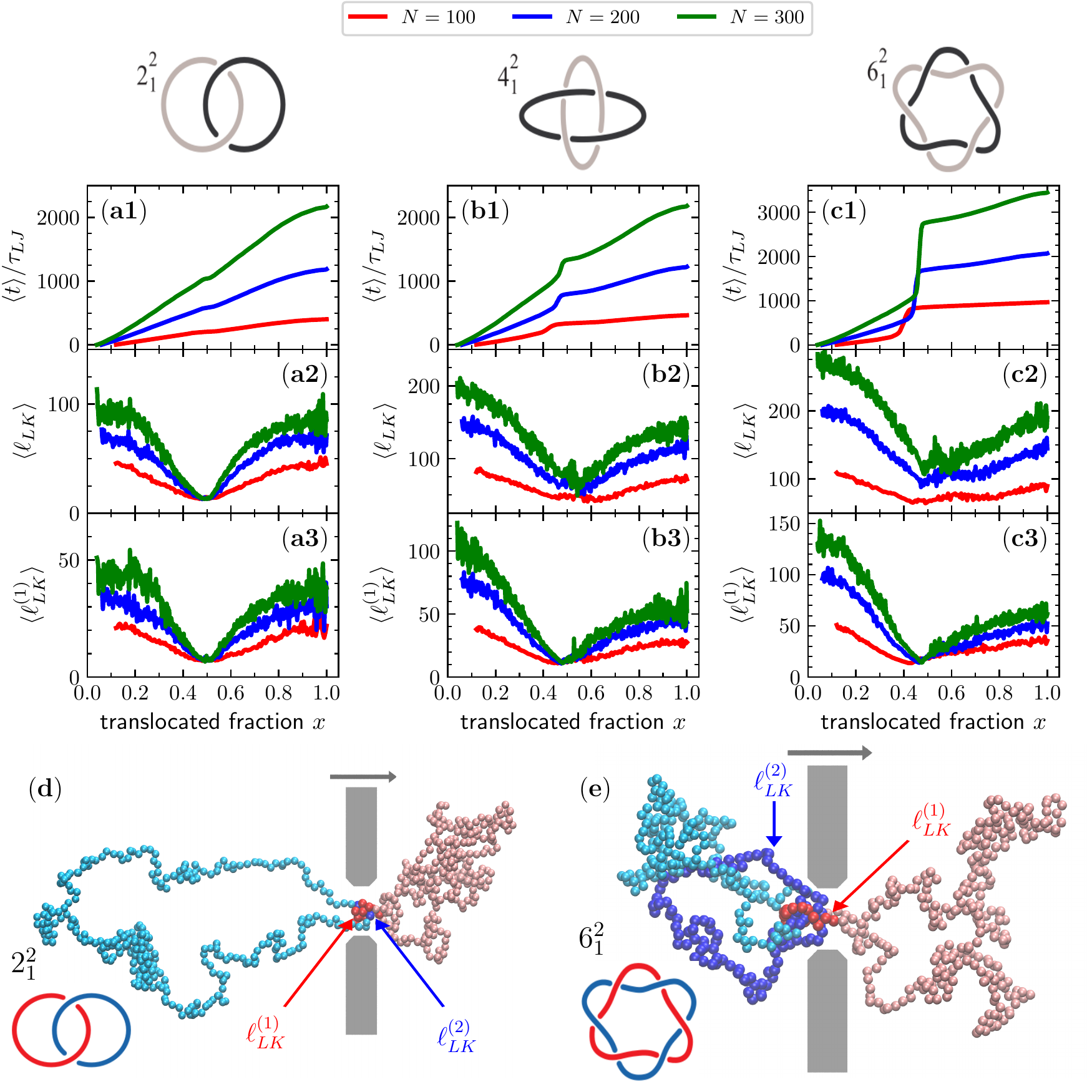}
 \caption{ 
 (a1,b1,c1) Average time required to translocate a fraction $x$ of respectively the
 Hopf link ($2^2_1$), the Solomon link ($4^2_1$) and the Star of David link ($6^2_1$).
 (a2,b2,c2) Average linked portion, $\ell_{LK}$, as a function of the translocated
 fraction $x$ for respectively $2^2_1$, $4^2_1$ and $6^2_1$.
 (a3,b3,c3) For the component which translocates first, average contour length
 participating to the linked portion 
 $\ell_{LK}^{(1)}$ ($\ell_{LK} =  \ell_{LK}^{(1)}+\ell_{LK}^{(2)}$), as a function of
 the translocated fraction $x$.
 In each panel, different curves refer to different values of $N$ according to the
 legend.
 (d) When $x \simeq 0.5$ both the portions $\ell_{LK}^{(1)}$ and $\ell_{LK}^{(2)}$ of
 the Hopf link  ($2^2_1$) giving rise to the physical link are small and independent
 of the total contour length of the link.
 (e) For the Star of David link  ($6^2_1$), at $x \simeq 0.5$, $\ell_{LK}^{(1)}$ is
 still small and independent of the total contour length of the link while the same
 does not hold for $\ell_{LK}^{(2)}$.
}
 \label{fig:Ndep}
\end{figure*}

We now address the issue of how the translocation of linked pairs of rings depends on their total size $2N$, \emph{i.e.} the total number of edges or bonds in the linked system.
In Figure~\ref{fig:Ndep}(a1-c1) we show the $N$-dependence of $\tau$ for the $(2,2k)$-torus links $2_1^2$, $4_1^2$ and $6_1^2$. 
As expected the total translocation time $\tau$ increases monotonically with $N$ for each link type; this is a simple extensive  effect that should hold also for a single unknotted ring.  
More interesting is the rate of increase of $\tau$ that depends on link complexity.
This gives rise to a peak in the waiting time curves located at $x \sim1/2$ and whose height, at fixed link type, increases with $N$ (not shown).
The fact that more complex and longer links hinder more markedly the translocation process at $x\sim 1/2$ should depend on how the physically linked region spreads along the system during translocation. 
We recall that  the linked portion is identified as the shortest portion of the two rings that, upon closure, has the same topology as the entire link (for a detailed description of the algorithm used to measure it see~\cite{Caraglio_et_al_ScieRep_2017,Caraglio_et_al_Polymers_2017}).
Denote by $\ell_K^{(1)}$ and $\ell_K^{(2)}$ the size of  the subchain respectively in loop 1 and 2 whose union defines the linked portion.
Its size is given by $\ell_K = \ell_K^{(1)}+\ell_K^{(2)}$.

In Figure~\ref{fig:Ndep}(a2-c2) we report $\langle \ell_K \rangle$ as a function of the translocated fraction $x$ for different link types and link length $N$.
One notices that at $x\sim 0 $ and $x\sim 1$, i.e. when the link does not experience the pore, the size of the linked region grows with $N$.
As the link is entering the pore the linked region shrinks reaching its minimum at $x\sim 1/2$ where the waiting time is maximum.
For the simplest Hopf link this minimum is independent on $N$ suggesting that, at $x\sim 1/2$, the linked region is localized within the pore (see Figure~\ref{fig:Ndep}(d)).
As link complexity increases, however, this minimum increases with $N$.
This is readily seen for the $6_1^2$ link in Figure~\ref{fig:Ndep}(c2). 
To understand this marked difference in the $N$ dependence of $\ell_K$ at $x\sim 1/2$  we have inspected more closely the configurations of $6_1^2$ as they translocate.
In addition to cases similar to the one observed for the Hopf links (localized linked portion) there are trajectories characterized by a strong asymmetry in the partitioning of the linked portion between the two rings (see Figure~\ref{fig:Ndep}(e)).
In particular one can notice that, while the essential crossings are still localized inside the pore, the subchain of the linked portion belonging to the second component is free to fluctuate in the CIS region.
This delocalised  variant of the link translocation should be more frequent for more complex links and we claim that it is responsible of the $N$ dependence of $\langle \ell_K \rangle$.
Indeed, if we extract from the total linked portion only the contribution coming from the beads belonging to the chain that translocates first, the $N$ independence at $x\sim 1/2$  is re-established also for the $6_1^2$ link.
It is worth noticing that a similar phenomenon has been observed in knot translocation~\cite{Suma_Micheletti_PNAS_2017}.

\section{\label{sec:level5} Discussion }

We have investigated the forced translocation of linked pairs and triples of polymer rings through a pore using Langevin dynamics.
For 2-component links with both rings unknotted the translocation rate depends on the link type.
If we plot the translocation time $\tau$ against the crossing number we obtain a set of relatively smooth curves for different families of links.
The cases that we considered were all 2-bridge links and it is convenient to represent
them as 4-plats.
The links in a particular family can differ in the number of intra-component or inter-component crossings and we show that these have different effects on the translocation process.
We notice that, for more complex links  the translocation time may depend on the order of entry of the components. 
This intriguing behaviour is observed only for some families of links and its full understanding requires more extensive and detailed studies.

If we consider 2-component links with one knotted component there are two general types.
The link can be prime or it can be composite, \emph{i.e.} the connect sum of a prime link with unknotted components and a knot.
In the latter case there are important differences depending on which component enters the pore first.
When the knotted ring enters first all of the crossings pass through more or less together and there is a single delay in the process.  
If the unknotted ring enters first there are two delays corresponding to the passage
of the crossings in the prime link and the crossings in the knot.
That is, the crossings in the knot enter the pore near the end of the process.
These findings suggest that translocation through nanopores can be exploited to distinguish between links and knots in pairs or rings.

For 3-component links we again have prime and composite cases.
For the prime cases like $6_1^2$ the crossings give rise to one delay, when the translocation is close to being one third completed.
For composite cases like the connect sum of two Hopf links, there are two delays, when the translocation is about one third and two thirds completed.
This could be used to distinguish prime and composite 3-component links experimentally.

The formulation of a theory of pore-driven polymer translocation is a highly non-trivial task that, in the past two decades, has been addressed for fully flexible and semi-flexible chains~\cite{Kantor2004,Sakaue2007,Sakaue2010,Rowghanian2011,Ikonen2012,Saito2012,Sarabadani2014,Sarabadani2017,Sarabadani2018}.
In particular, the Iso-Flux Tension Propagation (IFTP) theory~\cite{Sarabadani2014,Sarabadani2018} has been used successfully to describe results from MD studies and experiments in a variety of driven polymer translocation scenarios and has been generalized to include the case of two segments of a folded polymer passing simultaneously through the pore~\cite{Ghosh2020}.  This is an important starting point for dealing with knotted or linked polymers.
Clearly entanglement adds a further level of complexity to the problem and makes the development of a theoretical description of the complete translocation process in this case difficult, especially because the pore obstruction is dependent on the knot or link type and topology presumably plays a role in the propagation of the tension front on which IFTP is based.
However, Suma and Micheletti have shown that, when a knotted circular chain translocates through a sufficiently large pore, the pore obstruction caused by knot passage has a brief duration and the waiting time shows the typical behaviour predicted by IFTP with an initial phase corresponding to tension propagation and a second phase corresponding to tail retraction~\cite{Suma_Micheletti_PNAS_2017}.
In our study, in order to enhance the separation of responses depending on link type, the pore size is taken to be small enough that the pore obstruction due to entanglement strongly affects the translocation and waiting times.
Still, the waiting times in our simulations show a tension propagation phase behaviour for values of the translocated fraction smaller that the values at which the waiting time peak due to the pore obstruction due to entanglement arises (see e.g. Figure~\ref{fig:fig6}(b)).

Finally, it would be interesting to extend calculations of the type reported here either by changing the magnitude of the force or the size and geometry of the confining pore.

\subsection{Acknowledgments}
MC acknowledges financial support from the Austrian Science Fund (FWF): P28687-N27.

\section*{References}

\bibliography{References.bib}

\end{document}